\documentclass[onecolumn,showpacs]{revtex4}

\usepackage{graphicx}%
\usepackage{dcolumn}
\usepackage{amsmath}

\makeatletter
\def\btt#1{\texttt{\@backslashchar#1}}%
\DeclareRobustCommand\bblash{\btt{\@backslashchar}}%
\makeatother

\begin{document}


\title{Technology-Enabled Nurturing of Creativity and Innovation: A Specific Illustration from an Undergraduate
Engineering Physics Course}

\author{F.V. Kowalski, S.E. Kowalski, P.B. Kohl, and V.H. Kuo}
\affiliation{Physics Department, Colorado
School of Mines, Golden CO. 80401 U.S.A.}

\begin{abstract}

There is general agreement that creativity and innovation are desirable traits in the toolbox of 21\textsuperscript{st} century engineers, as well as in the future workforce in general. However, there is a dearth of exemplars, pedagogical models, or best practices to be implemented in undergraduate engineering education to develop and nurture those talents.
In this paper, we use a specific example of a classroom activity from a course designed to help bridge the transition from learning the fundamental principles of engineering physics in introductory courses to being able to creatively and innovatively apply them in more advanced settings, such as senior capstone projects and on-the-job challenges in the future workplace. Application of techniques for generating and evaluating ideas are described.
To enhance the benefits of group creativity and facilitate real-time electronic brainstorming in the classroom, we use InkSurvey with pen-enabled mobile computing devices (iPads, tablet PCs, Android devices, etc.). Using this free, web-based software in this setting effectively mitigates many of the social issues that typically plague brainstorming in a group setting. The focus, instead, is on paying attention to the ideas of others while encouraging fluency, originality, and honing positive critical thinking skills. This emphasis is reflected as the group creates a metric to evaluate their potential solutions.
A specific case from undergraduate and graduate level engineering physics courses is described to illustrate how the extensive work done in this arena in psychology, marketing, and business environments can be applied to STEM education. The classroom process is outlined and actual student results are presented to illustrate the method for other instructors who might be interested in employing similar activities in a non-threatening, low-stakes learning environment.

\end{abstract}


\maketitle

\section{Introduction}
\label{sec:intro}

In the landmark report about the Engineer of 2020, the first sentence in chapter one (p. 7) is: "Engineering is a profoundly creative process \cite{national}. Later in the report, creativity is targeted as one of a handful of essential qualities that are indispensable for engineers, growing in importance with the complexity of the challenges of the 21\textsuperscript{st} century (p.55).  Creativity manifests itself in science and engineering innovations, and few would disagree with the desirability of creativity and innovation in STEM graduates. Furthermore, a meta-analysis of 70 studies in a variety of disciplines unequivocally concluded that well-designed creativity training can be effective \cite{scott}.

The challenge for STEM educators, then, is in how those traits can best be nurtured and enhanced in the undergraduate curriculum.  A 2007 study supported the hypothesis that creativity is not valued in contemporary, mainstream engineering education, finding that the academic experiences of engineering students included almost none of ten factors identified for fostering creativity \cite{kazerounian}. A more recent study that used the previously validated Kirton Adaption Innovation Inventory and Abbreviated Torrance Test for Adults to assess the creativity of 200 first- and fourth-year engineering students found that ``though students' creativity may not be hindered by the current engineering curriculum, neither is their creativity enhanced" \cite{burgon}.

Looking at recent (2012) ASEE Conference proceedings \cite{web} to quickly take the pulse of the engineering education community, 13 papers and one workshop were strongly associated with how to best counter this creativity flatline. Although there is some overlap, these presentations fall into the following broad categories:
\begin{itemize}
\item 4 papers deal with creativity in the entrepreneur/business mindset;
\item 2 papers and one workshop concern creativity in the context of product design;
\item	2 papers describe trying to incorporate a creativity component in experiential and project-based learning;
\item	2 papers focus on general puzzles, activities, and games that are not engineering-specific, in efforts to stimulate creativity;
\item	3 papers deal with creativity and its role in engineering education in more general terms.
\end{itemize}
Considering this body of work, three areas of pressing need emerge.  First, as one of these papers specifically points out, there is a lack of research and resources on incorporating creativity in non-design engineering courses and those not focused on entrepreneurship \cite{zappe}. Second, many of these papers note an emphasis on group activities and a large fraction report difficulties associated with this. Since scientists and engineers often do not have the luxury of choosing those with whom they must work, a pedagogical model for enhancing group creativity would thus be useful.
Finally, in general,  there remains a gap between generic puzzles and games, which may be appealing to STEM students but do not require subject-specific knowledge, and more advanced requirements in the design or senior capstone context for students to demonstrate creativity.   With no intermediate steps, expecting a student to be creative and innovative on a design project, for example, is a bit like asking someone who cannot do a pull-up to hang on the bar until they can, instead of providing them with exercises to strengthen the specific muscles necessary achieve that goal. There is a need, then, for more specific examples of how educators can build up the skills associated with creativity in a subject-specific context.

A method to nurture creativity and innovation (C/I), useful in STEM education, is described here. We use a specific example of a classroom activity from a course designed to help bridge the transition from learning the fundamental principles of engineering physics in introductory courses to being able to creatively and innovatively apply them in more advanced (and less academic) settings, such as senior capstone projects and on-the-job challenges in the future workplace. While this application is in a classroom setting, its focus on enhancing group creativity has potential utility in workplace group interactions.

\section{The Complexity of Creativity}
\label{sec:complexity}

Creativity is a cognitive process involving both the retrieval of information and making novel associations and connections about this information. If the creative product is to be useful, these connections need to be critically evaluated to inform revisions. This leads to a C/I cycle where ideation is followed by critical thinking (involving both positive and negative aspects of the ideas). Of the two critical thinking components, the negative part is common but needs practice when applying it to content students have recently learned, particularly in STEM. Positive critical thinking, i.e., finding the positive aspects of an idea which might be disguised with many dysfunctional parts, is much less common. Both instructors and students can become more proficient at this with awareness and practice.

C/I can be accomplished individually or in a group environment, but focusing only on enhancing individual skills and traits ignores the complexity of C/I. Working in a diverse group can increase C/I \cite{milliken}.  Perhaps this is because a trait associated with creativity that is valued in one setting may not be advantageous in another.  For example, persistence in seeking new approaches becomes a liability if a product deadline is to be met, while it is an asset in situations with few time constraints. Questioning is an asset in challenging heuristics or models but is often a liability if applied in an unconstrained manner. Thus the collective creativity of a group can benefit from diverse traits brought to the table by individuals as they tackle problems. In addition, the literature suggests groups can enhance creativity by:  having ideas from different members of a group facilitate the retrieval of related ideas \cite{paulus1};  comparing different ideas, which induces a sense of competition \cite{coskun,paulus2}; exposing the ideas of others, which allows novel connections or combinations of existing ideas to be made \cite{nijstad};  and using the higher rate of idea generation to lead to more persistence \cite{nijstad}.

Nurturing group C/I is of particular importance in STEM education for a variety of reasons. Since classes are usually taught as groups, this is an easy, accessible, and familiar environment. Like many things in the classroom, even though techniques for enhancing C/I can be practiced in a group, they can also be used individually.  Moreover, complex problems often are beyond the capacity of an individual and require group solutions.  Mirroring the workplace environment, the problems addressed can be open-ended and their resolution involves social interaction.

However, groups can also discourage creativity; others have identified four primary ways in which this can happen. In a group setting, there may be ``evaluation apprehension," a fear about how one's ideas are judged by others \cite{paulus1}. Also, there may be ``production blocking," or interference in one's ideation while others exchange ideas \cite{diehl}. There may be reduced accountability in a group setting and therefore performance degradation \cite{karau}.  Finally, generation of common or similar ideas may make it difficult for the group to move forward \cite{nijstad,paulus3}.

Simply expressing the positive and negative aspects of the C/I process in a group setting, though, does not reflect its complexity.  For example, exposure to the ideas of others (listed as a positive variable above) can also lead to a fixation effect, reducing the novelty of ideas generated by the group \cite{kohn1}. There also may be complex interactions between the many factors influencing C/I. A group member strongly motivated to solve a problem, for example, might be less affected by evaluation apprehension. Thus what may appear as an independent variable could turn out to be influenced by other factors.

Nevertheless, the strength of the reasons given above for enhancing group C/I motivates efforts to enhance the positive while mitigating the negative effects of the group environment; technology provides a promising avenue.

\section{Using Technology in the Training Cycle}
\label{sec:trainingcycle}

A simple technique for teaching creativity and innovation is difficult to construct due to this complexity of the creative process. Teaching any skill, however, typically involves an iterative procedure, where students are shown an example, followed by practice, and then assessment with feedback. This is referred to as a training cycle. Depending on the outcome of the assessment, the cycle is repeated until a high success rate is achieved.  Such a strategy has been successful in training for skills and content knowledge. Nurturing C/I in students is even more complex, since it involves additional skills (such as listening, problem solving, being observant, and noticing anomalies), along with both traits and habits (such as being curious, seeking patterns, remembering, being persistent, and paying attention to the ideas of others) and motivation (extrinsic and/or intrinsic).

Collecting real-time assessment during the iterative procedure might not be effective in a classroom due to students' concern about how their ideas are judged by others (evaluation apprehension, see above). One solution which mitigates this issue while maintaining the advantages of group interaction is to use technology to transmit student responses in real time to the instructor, who can then share selected responses anonymously with the class. Students can be equipped with various devices to submit typed responses, or with pen-enabled Tablet PCs, iPads, and Android devices to use digital ink to construct sketched, graphed, diagrammed, etc. responses. While there are a variety of software products that perform this function, {\em InkSurvey} was used in this study and it will be used as a generic descriptor for a product with this type of functionality.  On a web page, each student enters sketched and/or typed responses to open-format questions posed by the instructor. {\em InkSurvey} works well with large numbers of respondents, is designed to transfer digital ink responses, is web-based and platform independent, and is free \cite{kowalski}.

Student responses can be viewed instantaneously on the instructor's web page. The submitted responses can then be selectively and anonymously shared with the rest of the class and serve as effective springboards for subsequent class discussions. As in any skills training, students learn to be more fluent by watching the instructor and fellow students participate, as well as through scaffolded instruction from the instructor. This use of technology is useful in the training cycle since it actively engages the students and enhances metacognition by requiring students to write about their understanding through open- format questioning, rather than by selecting responses from a multiple choice menu. In addition to these advantages, the applicability here to facilitating electronic brainstorming is noteworthy.

\section{Electronic Brainstorming}
\label{sec:brainstorming}

Rules for group ideation, called brainstorming, were first developed by advertising executive Alex F. Osborn \cite{osborn}. His procedure attempts to generate a large number of ideas, many of which are impractical \cite{sappington}. Constraints in science and engineering, however, require a practical solution. Therefore the critical thinking component of C/I probably plays a more important role in STEM brainstorming than in the arts and business.

Some studies of the efficacy of brainstorming suggest that nominal groups with no social interaction outperform interacting groups \cite{goldenberg}. The metric however, typically involves quantity of ideas rather than quality, which is much more difficult to judge. There is also concern that the subjects and problems addressed in these studies do not reflect performance in a workplace environment \cite{domburg}. Another concern is that brainstorming groups often do not choose novel ideas but rather what they think is feasible \cite{kohn2,putman,rietzschel}.

The use of computer technology to enhance group work by allowing participants to exchange ideas and comments is referred to as a group support system (GSS)\cite{dennis1} or an electronic meeting system (EMS). When the technology is focused on enhancing ideation, this often becomes electronic brainstorming (EBS) \cite{dennis2}. In one of the first uses of EBS, 16 group members worked on individual computers to enter and view anonymously all ideas from the group \cite{nunamaker}. A review of other systems is given by Procter \cite{proctor}. More recent efforts are by Hermann and Nolte \cite{hermann} and Kakoulli \cite{kakoulli}, who records the meeting. The last two papers give details of current EBS techniques, while Kakoulli also has a comprehensive discussion of ideation methods.

\section{Process for Nurturing Creativity}
\label{sec:process}

Based on the current understanding of creativity, the first iteration of our process for nurturing creativity begins with statement by the instructor of a problem or goal, and then involves requests for: (1) ideas, (2) questions, (3) positive critical comments on the ideas, (4) negative critical comments on the ideas, (5) construction of a metric to determine the solution based on these positive and negative comments, and (6) new ideas after practicing methods for idea generation. This procedure is cycled to refine the solution.  Depending on the complexity of the problem, the entire process could extend over multiple class sessions.  A more detailed description of each step is now given.

First, students are asked to individually and anonymously (to their peers but not the instructor) submit unrestrained ideas (not just ideas that students think are practical) using digital ink (pen-enabled mobile devices such as tablets, iPads, and Android devices) and {\em InkSurvey}. This is done without a discussion with other group members. During the collection of these ideas, input on any questions the students have can simultaneously be submitted. The questions may address constraints in the problem, factual information, motivation for studying the problem, etc. As the questions are received, they are discussed by the group, which often leads to the generation and submission of additional ideas.  The facilitator organizes the ideas and displays them to the group, triggering discussion by the group as appropriate. The ideation request can then be repeated if appropriate.

Using the organized list of ideas, the facilitator then asks students to use {\em InkSurvey} to make positive comments about these ideas. This forces the students to pay attention to the ideas of others and engage in positive criticism. It may be necessary for the facilitator to provide an illustration of positive criticism, as many students are not accustomed to seeking the ``diamond in the rough" in the ideas of others.  Again, the facilitator organizes these comments as the procedure is repeated for negative critical thinking comments. The organized critical comments are then presented to the group and discussed as appropriate. The positive and negative comments then constitute the foundation for constructing the metric that will guide selection of the final group solution.

Finally, this process is repeated to generate and refine additional ideas. In these subsequent iterations, various techniques for generating new ideation may be productive; some standard techniques include:
\begin{itemize}
\item Using a random word as a prompt to break a fixed thought pattern/heuristic/model and generate novel connections \cite{debono}. For example, students could be asked to come up with science/engineering terms and then use these words as priming in a repeated application of the ideation question.
\item	Asking for questions of clarification, motivation, needed factual information, or simply repeated application of ``why," which often reveals tacit assumptions.
\item	Having students sort the group ideas into categories. This reveals deep-seated thought patterns which, when exposed, can lead to breaking these patterns. As these patterns are displayed to the group, the facilitator can point out how fluency and originality match a particular categorization. Fluency in each category can then more easily be extended in new ideation. This is also an example of additional practice in paying attention to the ideas of others.  Each person may create a different set of categories; the group can then critique these categories.
\item	Asking for analogies. For example, Kepler used an analogy with the intensity of light decreasing from a point source to infer how gravitational attraction decreases with distance; Mueller used an analogy of transmuting atoms with radiation to transmuting genes with radiation.
\item Having each student sketch a mind map and display it to the class.
\item Showing pictures and/or videos related to the problem to stimulate associations and remembering.
\item Asking students to focus on anomalies or behavior which doesn't match the accepted pattern/heuristic/model for the problem.
\end{itemize}

\section{Specific Example from Engineering Physics Classrooms}
\label{sec:example}

This process for nurturing creativity and innovation has been administered separately to two classes: a sophomore class of 23 engineering physics students (who had completed 2 semesters of physics, engineering design, and chemistry along with three semesters of calculus), and a class of 7 graduate physics students. The classes met one hour a week in an interactive lecture format. Each student used a tablet computer in class to submit responses to the open-format questions via {\em InkSurvey}. To specifically illustrate both the methodology described above and examples of how the students responded to it, we next discuss one project used in the classes; results from both classes are combined here.\\*\\*
\underline{The Problem}: Students were asked to generate ideas on how to measure the mass removed from a quarry.  No additional information was given.\\*\\*
\underline{Initial ideas} submitted by students were to measure:
\begin{enumerate}
\item the weight of loaded and unloaded trucks as they carried the material out of the quarry and returned empty, with the difference showing the mass removed;
\item the volume of rock extracted, and then calculate the mass from the rock's density;
\item   the flow rate of material on a conveyor belt loading the trucks;
\item the fuel consumed by machinery moving the rock, and then calculate the mass from the energy required to move it;
\item the mass, via a scale incorporated into the loader, as the rock is loaded onto the truck;
\item the displacement of the mass in a fluid, to calculate the volume and then the mass using the density of the rock;
\item the volume removed, using laser imaging of the quarry;
\item the oscillation frequency of removed mass placed on a stiff spring;
\item the strain in the arm of a machine moving the mass;
\item the volume of liquid rock after it has been melted;
\item the terminal velocity of extracted rock as it flows through a fluid; and
\item the effect of mass loss on the gravitational attraction at some point in the quarry.
\end{enumerate}

The request for questions yielded the following results:\\*
Are we looking for a generic solution or one for a particular material?\\*
What methods are used to move rock?\\*
How does rock density vary?\\*
What power methods are used to move rocks?\\*
What methods are used to extract rock?\\*
What methods are used to measure volume, weight, power?\\*
How much energy is needed to melt rock?\\*
How small of a gravitational attraction can be measured?\\*
What changes when rock is moved/removed?\\*
How can we detect that change?\\*
What limitations arise based on the material being removed from the quarry?\\*
Why are we interested in this problem?\\*

Answers to these questions were then discussed in class. Before repeating the request for ideas, students searched the web for images/videos of quarries.\\*

Next, a request for \underline{positive critical thinking comments} about the collected ideas resulted in these remarks (numbers correlate to list of initial ideas):\\*
(\# 3)  for measuring the flow rate of rock on a conveyor belt: speedy, measures and moves at the same time, cost effective since you don't need another instrument;\\*
(\# 4)  for measuring the gas in the truck: is free of other equipment, accurate if there is constant speed on a flat highway;\\*
(\# 5)  for measuring the mass with the loader: is efficient, since the mass has to be moved anyway, it is less expensive than a large scale on the truck, just weighs the rock and not rock plus truck;\\*
(\# 8)  for measuring the oscillation frequency after putting the rock on a spring: implementation straightforward, frequency measurement is accurate;\\*
(\# 12) measuring the gravitational attraction is very accurate.\\*

\underline{Negative critical thinking comments} included:\\*
(\# 3) measuring flow rate on a conveyor belt: hard to measure the flow accurately, not a continuous measurement;\\*
(\# 4) fuel measurement in the truck: consumption depends on engine wear, temperature;\\*
(\# 8) measuring the oscillation frequency after putting the rock on a spring: spring deterioration might be a problem, time consuming, environmental wear, limited amount can measured, hazardous to humans, spring calibration needed; and\\*
(\# 12) measuring the gravitational attraction is impractical to implement.\\*

The \underline{metric} to evaluate the final solution, deduced from both positive and negative critical thinking, was agreed upon to include the following considerations: (1) the extraction, measuring, and transport should be at the same time, (2) the solution should use something already in place (fuel, conveyer belt, etc.), (3) the method should measure the mass multiple times at different steps in the process, (4) the solution should be non-intrusive, (5) the technique should be quick, not slowing the extraction process down, and (6) the method should be useful in multiple quarries.

Next, techniques for generating \underline{new alternative ideas} were practiced as described above. We report here only on the technique to classify the group ideas, with the following results:
\begin{enumerate}
\item Classification Scheme A. One classification system the group devised consisted of the following categories:
    \begin{itemize}
 \item Ideas that use a scale to weigh the rock. Fluency involves weighing a truck, weighing the machine which picks up the rock, etc, (ideas \# 1, 5, and 9);	
\item Ideas that use fuel consumed in moving the rock. Fluency involves measuring the gas used by a truck over a known trajectory, the electricity for an electric engine, etc. (idea \# 4);
\item Ideas that measure the volume removed. Fluency involves using a laser ranging system to map the quarry before and after, using the displaced volume of water if the rock moves on a conveyer belt through water, etc. (ideas \# 2, 6, 7, and 10); and
 \item Ideas that measure the flow rate of rock out of the quarry. Fluency involves using a conveyor belt and monitoring the volume of rock that flows, measuring the flow rate on some slide that the rock moves down, etc. (idea \# 3, 11).
\end{itemize}
\item Classification Scheme B included all ideas measuring volume. Fluency involves using the volume of rock, volume of gas used by the truck, or volume of rock removed to measure the mass removed.
\item Classification Scheme C.  This group also saw a pattern with these three categories:
    \begin{itemize}
 \item ideas that involved weighing with a scale (ideas \# 1,5, and 9);
\item ideas that measure the volume displacement and then use density to find mass (ideas \# 2, 6, 7, and 10); and
\item ideas that measure the flow rate of any substance, including rock and fuel (ideas \# 3, 4 and 11).
    \end{itemize}
\end{enumerate}

Next students were asked if these classifications generated new ideation or if these classifications could be combined. For example, in the laser volume measuring idea (\# 7), the location of the ranging device varied (tower, plane, on the extracting machine) and the radiation source varied (acoustic waves, lasers, ultraviolet radiation, neutrons, radio waves, microwaves). Combinations of these were discussed. An example that resulted is a device that measures the volume of rock moved on a conveyer belt or in a truck bed using acoustic waves, since dust scatters most other radiation.

A matrix was also used to generate new idea combinations.  The columns showed extraction methods (truck, elevator, conveyer belt, and loader) and rows indicated energy conversion mechanisms (gas, electric, pneumatic, hydraulic). One new idea that emerged was that of using the electrical power consumed by a conveyer belt to measure the extracted mass.

At this point, the group chose two methods of measuring mass: by using acoustic radiation to measure its volume as it moved on a conveyer belt or rested in a truck bed, and by monitoring the electrical power consumed by the conveyer belt. Repeating the critical thinking skills question resulted in practical questions about implementing the first method. While these questions were difficult to answer for the first method, on-line calculators for electrical power consumption in a conveyer belt were readily available and directly related to implementation of the second method.

The decision metric developed earlier for this problem (see above) guided the class to choose calibrating the power consumed by a conveyer belt as the method of determining the mass removed from a quarry, since all of the points in the metric were satisfied. The volumetric measurement, on the other hand, seemed to require new technology and thus fell short on satisfying the metric.

\section{Discussion}
\label{sec:discussion}

This procedure for nurturing C/I in a STEM-specific environment has many demonstrated advantages.  Every student is actively engaged in the process. In addition to explicitly nurturing creativity in the students, it also refines their critical thinking skills as they evaluate their own ideas and those of others to agree on a decision metric.

Throughout the process, time is devoted to development of both individual skills (through ideation and critical thinking) and group collaboration (through brainstorming and generation of a metric). After collecting ideation individually, the facilitator organizes the ideas before presenting them to the group for analysis, thus combining individual and group C/I and also forcing the student to pay attention to the ideas of others. This process better equips students with tools to creatively and innovatively apply their understanding of fundamental principles in STEM to larger, more complex problems in future courses and the workplace.

Part of our approach to constructively deal with the social issues of the group is the use of {\em InkSurvey}. Through the use of readily available technology, input from all the students can be received instantaneously and organized for further class discussion. This addresses evaluation apprehension, since student responses can be shared anonymously with the audience and no one needs to fear embarrassment because of their response. The anonymity of the responses also allows unbiased evaluation of all ideas, regardless of the gender, race, or social status of the student source. The instructor, however, can view who has submitted each response, which establishes individual accountability and mitigates the group performance degradation issue discussed above.  Requests for ideas, questions, and feedback during the cycle also enhance metacognition by requiring students to write about their thoughts and what they do and do not understand.
In using this procedure, the positive aspects of group interaction not only remain, but are nurtured. For example, ideas from different members of a group are not only shared but attention to these ideas is facilitated. Ideas are compared, generating a sense of competition. Novel connections or combinations of existing ideas are stimulated as the cycle completes and repeats. Persistence is encouraged both by generating new ideas and by allowing the class to anonymously express any motivational concerns which are then addressed by the facilitator.

As discussed earlier, there is a concern in the dynamics of group creativity that there may be difficulty in the group moving forward if they generate common or similar ideas.  However, the menu of techniques for generating new ideas has been fruitful in circumventing this potential negative group interaction. Likewise, the concern that groups often do not choose novel ideas but rather what they think is feasible (see earlier) is addressed here by having the group develop a metric at an early stage and using it to guide the final decision.

The utility of this work rests in part on using a problem that does not have a single solution. Indeed, a realistic evaluation depends on many issues not discussed in the context of this classroom example, such as cost, reliability, design life, safety, and environmental impact. Nevertheless, this exercise illustrates many of the fundamental pitfalls and advantages in dealing with group creativity and how to overcome them; the product of this process could then be further analyzed in light of some of these more pragmatic issues.

From the perspective of the instructor, perhaps one of the most enlightening and interesting aspects of this work is how difficult it is to avoid making judgments about the ideas and comments generated, particularly those dealing with positive and negative critical thinking. For example, consider the positive critical comment that using a spring to measure the mass extracted is both easy and accurate. This would most likely never be brought forward by an expert, who has internalized heuristics/models of machinery showing this to be impractical. Nevertheless, the idea was not immediately discarded and after applying the techniques for generating alternative ideas, this gave rise to the concept of measuring the vibration frequency of oscillations of a conveyer belt mechanism, which seemed to be worth further investigation. Similarly, the use of lasers to measure the volume of extracted rock in a dust-filled environment seemed naive due to the amount of light scattering. However, applying the above procedure transformed this idea into a more realistic one, that of using acoustic waves (which scatter little in dust composed of particulates whose size is less than the acoustic wavelength) to measure the volume of rock moving on a conveyer belt.

This illustrates why it is so difficult to be creative in STEM. The novel associations and connections that we make about information are hidden by our heuristics/models of how the world works. A child who questions everything has yet to develop these heuristics/models. As they are developed, new information is framed in this context. The process described here attempts to break out of the many constraints encountered in STEM while creating a non-threatening, low-stakes learning environment.  In the classroom, it helped both students and instructors evaluate their existing heuristics/models and rewarded them with new insights.  Since most STEM students are unfamiliar with this type of pedagogy, it is critical that the instructor move away from the traditional ``right or wrong" evaluation techniques.  In turn, students are made aware of and become comfortable with the notion that they are free to make errors without any consequences, and that it is active participation that increases their learning and adds to their skills. The procedure described in this paper allows for exactly that.

It is worth noting that the role of technology in this process is simply to facilitate the exchange of ideas and comments in a real-time, anonymous environment.  Because all student input is processed by the instructor, the strength of this method to nurture the C/I process is therefore not inherent in the software, but rather in the skill of the instructor in filtering and responding to student input and guiding subsequent class discussions. Further investigation is needed to explore the successful transferability of this method to other instructors.  In some cases, particularly with larger classes, some of the instructor's organizational load could be postponed until less hectic times between class meetings; in our experience, this does not seem to have negative implications for the value of the method, and in some cases has enhanced it.

Other issues about the utility of this method remain. It suggests a host of research questions worthy of further exploration, including: Can other components of the C/I process (curiosity for example) be taught independently, and if so which ones? How does training in these individual components affect the complete C/I process? In how large of a group could this C/I cycle be used effectively?

\section{Conclusions}
\label{sec:conclusions}

This paper describes and provides a classroom example of a process designed to nurture creativity and innovation in students of engineering and other STEM disciplines. A fundamental aspect of this process is using pen-enabled mobile technology to facilitate the exchange of ideas, thus allowing both individual and group creativity to be developed and enhanced.

{\em InkSurvey} is free, web-based software that allows students to use keyed input or digital ink to construct open-format responses, which are received by the facilitator or instructor instantaneously.  Thoughts submitted in this manner can then be anonymously displayed to the entire class.  This is particularly useful in enhancing the creative process of a group since it reduces student apprehension, requires accountability, and encourages participation.  Additionally, this use of technology focuses the group on interaction through communication and careful examination of the ideas of others.

The unique contributions of the process described here include its focus on a classroom with large numbers of students, its utilization of mobile technology, its emphasis on nurturing C/I skills in a non-threatening, low-stakes, STEM-specific learning environment that is neither product design- nor entrepreneurially-focused, its generation of a decision metric from positive and negative critical thinking of the group, and its incorporation of both individual and group ideation.

\begin{acknowledgments}
We are very grateful for support received for this work from the HP Catalyst Initiative; this includes the Tablet PCs used in the classroom as well as support for the development and accessibility of {\em InkSurvey}.  In the design and delivery of the engineering physics class on nurturing creativity, we greatly appreciate the help and encouragement of Dan Leach and John Trefny.
\end{acknowledgments}

\end{document}